# Stability and Dynamic of strain mediated Adatom Superlattices on Cu<111>

W.Kappus

wolfgang.kappus@t-online.de

v01: 2012-05-11

## Abstract

Substrate strain mediated adatom density distributions have been calculated for Cu<111> surfaces. Complemented by Monte Carlo calculations a hexagonal close packaged adatom superlattice in a coverage range up to 0.045 ML is derived. Conditions for the stability of the superlattice against nucleation and degradation are analyzed using simple neighborhood models. Such models are also used to investigate the dynamic of adatoms within their superlattice neighborhood. Collective modes of adatom diffusion are proposed from the analogy with bulk lattice dynamics and methods for measurement are suggested. The explanation of surface state mediated interactions for superstructures found in scanning tunnelling microscopy experiments is put in question.

## 1.Introduction

Adatom superstructures have gained much interest for general and technological reasons, recent reviews were given in [1,2]. The general interest comes from the search for lateral interactions on surfaces reponsible for self organization of adatoms and related consequences in surface physics and chemistry. The technical interest comes from the hope for new methods in nanoclustering to enable new devices based on quantum effects.

Adatom superlattices on Cu<111> have been a subject of intense recent research [3,4,5] and the scanning tunneling microscopy experiments provide challenges for their explanation and further theoretical work. Superlattices are stable adatom structures with a long range order and lattice constants significantly larger than the lattice constant of the substrate. Key topic hereby are the lateral interactions by which the superlattices are mediated and stabilized. Though one explanation - surface state interactions - currently dominates we will put that in question below and will offer and justify an alternative - substrate stress mediated interactions.

Substrate stress mediated adatom interactions are not at all a new topic but were reemphasized recently in conjunction with a Born-Green-Yvon based theory of the adatom pair distribution [6]. The lack of experimental data made this paper somewhat speculative, this was the reason to search for a possible application case.

In this work, based on [6,7,8 ], the structure, stability and dynamic of adatom superlattices on Cu<111> mediated by substrate strain is outlined and discussed. In section 2 the model details are recalled and slightly modified against [6]. In section 3 the results of the adatom density distribution calculation and of a complementing Monte Carlo simulation are presented. In section 4 conditions for stability of the superstruc-ture against nucleation and melting and the dynamic of adatoms in their superstructure are discussed. In section 5 the current explanations for experiments are challenged. In section 6 results and consequences are discussed and methods for further verification are sketched. Also open questions are adressed. Section 7 closes with a summary of the results.

## 2. Model Details

- 2.1. Adatom Density Distribution

The relation between the pair distribution $g(\vec{s}_1,\vec{s}_2) \equiv g_{12}$ of adatoms on a surface, the temperature scaled adatom-adatom interaction $u_{12}=U_{12}/k_BT$, and the adatom coverage $\theta$ in Kirkwoods superposition approximation is given by an adapted Born-Green-Yvon equation [6]

$$\ln g_{12} + u_{12} = (2\pi)^{-1} \theta \int_S [g_{23}-1] d\vec{s}_3 \int_S g_{43} |s_{14}|^{-2} \vec{s}_{14} \cdot \vec{\nabla}^{(4)} u_{43} d\vec{s}_4. \qquad (2.1)$$

Here $\vec{\nabla}^{(4)}$ acts on the coordinates of adatom (4) only. The left hand side of (2.1) describes the familiar zero coverage case while the right hand side of (2.1) introduces the coverage dependent part. This non linear equation for $g_{12}$ will be solved numerically below for the special case of $u_{12}$ describing strain mediated interactions on Cu<111>.

Computing $g(\vec{s}_1,\vec{s}_2)=g_{12}$ for different $\theta$ from (2.1) is straightworward starting iterations from zero coverage and increasing $\theta$ slowly. The iteration step leading from $g_n$ to $g_{n+1}$ for coverage $\theta_i$ is

$$g_{n+1} = \lambda \exp(-u + \theta_i F(g_n)) + (1-\lambda) g_n, \qquad (2.2)$$

where $\lambda$ denotes a damping parameter. $\lambda$ can be chosen 1 for small coverages $\theta_i$ but must be reduced towards 0.1 when $\theta_i$ approaches a critical coverage $\theta_c$.

- 2.2. Elastic Interactions of Adatoms

Following [6] the interaction of adatoms located at the origin and s using polar coordinates (s,$\phi$) for their distance $s = |\vec{s}|$ and pair direction angle $\phi$ with respect to the crystal axes is given by

$$U_p(s,\phi) = (2\pi)^{-1} \sum_p \omega_p \cos(p\phi) \cos\left(p\frac{\pi}{2}\right) \Gamma\left(\frac{p+3}{2}\right)$$

$$s^p \,_1F_1\left(\frac{p+3}{2}; p+1; \frac{-s^2}{4\alpha^2}\right) \Big/ \left(2^{p+1} \Gamma(p+1) \alpha^{p+3}\right), \qquad (2.3)$$

where $_1F_1$ denotes the Hypergeometric Function, $\Gamma(p)$ the Gamma function, $\alpha=\sqrt{2}/2$ is a cutoff length defining height and location of the potential wall and the medium range potential, and the $\omega_p$ denote coefficients of a cosine series [8]. The dominating isotropic p=0 term of (2.3) is negative for small s describing a potential well (i.e. an attractive potential), has a positive wall (i.e. a repulsive potential) at $s=s_w$ and approaches infinity with a $s^{-3}$ law. For anisotropic substrates like Cu the p>0 terms describe the anisotropic part of the interaction and influence the height of the positive wall in dependence of the pair direction angle $\phi$ with respect to the crystal axes. Tab. 1 shows the $\omega_p$ for the elastic adatom interaction on Cu<111> calculated as outlined in [6]. We note the units of the $\omega_p$:
- the numerator is $P^2$, the square of a scalar parameter P describing the lateral stress magnitude an adatom excerts to the surface
- the denominator is the $c_{44}$ elastic constant of the substrate.
For details of the parameter P see [6,8].

| Substrate | $c_{11}$ | $c_{12}$ | $c_{44}$ | $\zeta$ | $\omega_0$ | $\omega_6$ | $\omega_{12}$ |
|---|---|---|---|---|---|---|---|
| Cu | 169. | 122. | 75.3 | -1.376 | -1.01 | -0.007 | +0.0004 |

Table 1. Substrate Elastic Constants $c_{ik}$ (GPa) from [9], anisotropy $\zeta=(c_{11}-c_{12}-2c_{44})/c_{44}$ and coefficients $\omega_p$ (in $P^2/c_{44}$ units) on Cu <001>.

A cap

$$U(s, \phi) = U_w + U_{wp} \cos(p\phi) \frac{s}{s_w} \quad \text{for } s < s_w \quad (2.4)$$

is used to avoid the problem of an exploding pair distribution $g_{12}$. This trick allows to treat a non-equilibrium problem (created by nucleation) with equilibrium methods.

■ 2.3.Complementing Monte Carlo Simulations

The pair distribution function g(s) can be interpreted as statistical average over many adatom position samples. Monte Carlo simulations of adatoms interacting with the capped interaction have been performed. In accordance with the continuum model used for describing the interaction, a grid-less algorithm has been used. Periodic boundary conditions were applied. The area size of 60 units was chosen to keep the computing time in the range of hours while the interaction u(s=60) has decreased well below 0.001. Starting from a random k member adatom configuration $\{\vec{s}_{i,0}\}$, step n + 1 $\{\vec{s}_{i,n+1}\}$ evolves from step n $\{\vec{s}_{i,n}\}$ by

$$\vec{s}_{i,n+1} = \vec{s}_{i,n} + \chi \sum_{j=1}^{k} \vec{\nabla}^{(j)} u(\vec{s}_{i,n}, \vec{s}_{j,n}) + \vec{t}_j, \quad (2.5)$$

with an appropriate parameter $\chi$ and random displacements $\vec{t}_j$ to account for the thermal movement of the adatoms. So an adatom moves around under the forces of all its neighbors until all forces are balanced. Convergence is achieved up to the critical coverages $\theta_c$ for $\vec{t}_j=0$. For $\vec{t}_j>0$ the $<\vec{s}_i>$ converge in a coverage range $\theta<\theta_c$. Intentionally and in line with the continuum model used, also the random displacements are chosen off the substrate grid.

# 3.Calculations and Results

■ 3.1.Adatom Pair Distribution on Cu <111> Surfaces for increasing coverage

The presentation contains 360° contour plots with the reference adatom in the origin and value legends in the lower left quadrant. The contour plots range from s = 0 to 22 to emphasize the relevant region. Furthermore g(s, $\theta$) evolution plots are shown. Especially:
- Fig. 1.a shows an u(s) contour plot, dark colors represent high values, light colors represent low values,- contours changing in 1- steps. The elastic anisoptopy is small but remarkable.
- Fig. 1.b shows a g(s) contour plot for the limiting coverage $\theta_c$, dark colors represent high values, light colors represent low values, contours changing in 0.1 steps. Dominant peaks are at the next neighbor positions of a <1-10> aligned hexagonal close package with an average lattice distance of about 5 substrate lattice spacings. The minima at the second next neighbor positions indicates a trend towards voids in a denser lattice instead towards a perfect less dense lattice. The superlattice constant of a perfect hexagonal close package of coverage 0.045 ML would be 5.07.
- Fig. 1.c shows a semi-logarithmic $g_{<1-10>}(s,\theta)$ plot for s || <1-10>, showing its evolution with increasing coverage $\theta$, the dominant peak develops towards 5 spacings. Also oscillations develop with a length of ca. 5 spacings. At small distances g(s) is strongly increasing when the covarage approaches its limiting value incdicating the onset of nucleation.
- Fig. 1.d shows a semi-logarithmic $g_{<-1-12>}(s,\theta)$ plot for s || <-1-12>, showing its evolution with increasing coverage $\theta$, a less dominant peak develops towards 5 spacings.

Not surprisingly the figures look similar to those of Au<111> [6] due to a similar substrate anisotropy. The higher value of the limiting coverage $\theta_c$ of Cu<111> compared to Au<111> is due to a slower increase of $\theta_i$ in the (2.1) algorithm and therefore has almost no physical meaning. We note that also Ag has a similar anisotropy, so one would expect similar adatom pair distributions.

- ### 3.2. Adatom Pair Distrubution on Cu <111> Surfaces for increasing temperature

    To illustrate the degradation of the superlattice order with increasing temperature at a constant moderate coverage of $\theta$=0.024 g(s, T) evolution plots are shown. Temperature varies from T=1 to T=2.2 in $U/k_B u$ units.
    - Fig. 1.e shows a semi-logarithmic $g_{<1-10>}(s,\theta)$ plot for s ∥ <1-10>, showing its evolution with increasing temperature T. The peak is decreasing and the g values for small distances $s<s_w$ are increasing, an indication of nucleation and superlattice degradation.
    - Fig. 1.f shows a semi-logarithmic $g_{<-1-12>}(s,\theta)$ plot for s ∥ <-1-12>, showing its evolution with increasing temperature T.

- ### 3.3. Adatom Pair Distrubution on Cu <111> Surfaces for decreasing temperature

    To illustrate the stabilization of the superlattice order with decreasing temperature at a constant moderate coverage of $\theta$=0.024 g(s, T) evolution plots are shown. Temperature varies from T=1 to T= 0.44 in $U/k_B u$ units.
    - Fig. 1.g shows a semi-logarithmic $g_{<1-10>}(s,\theta)$ plot for s ∥ <1-10>, showing its evolution with decreasing temperature T. The peak is increasing and shifting to larger distances indicating a more pronounced superlattice with a greater lattice constant.
    - Fig. 1.h shows a semi-logarithmic $g_{<-1-12>}(s,\theta)$ plot for s ∥ <-1-12>, showing its evolution with decreasing temperature T. The peak is nearly constant but is shifting to larger distances.

- ### 3.4. Adatom Superlattices on Cu<111> by Monte Carlo simulations

    The Monte Carlo simulations give an impression how the adatom pair distribution originates. Averaging over many samples would lead to a result as presented in 3.1, especially if thermal effects are accounted for. Samples show a strong tendency towards a substrate aligned hexagonal close package as expected from 3.1.

    Fig.1.i shows an adatom position sample from a Monte Carlo simulation according to (2.5) with zero random displacement at coverage $\theta$=0.045. Dimers or clustered objects with a distance $< s_w$ are depicted in red, others in blue color.

    We note the alignment towards <1-10> directions, a significant number of dimers, voids in the hexagonal close packed structure and a superlattice constant of about 5 substrate lattice spacings.

- ### 3.5. Adatom Movement within the Monte Carlo simulation

    In case the thermal random displacement $\vec{t}_j$ values in eq. (2.5) are chosen in the region {-1,+1} the stability of eq. (2.5) is limited to coverages $\theta < \theta_c$.

    Fig.1.j shows an adatom trajectory sample from a Monte Carlo simulation according to (2.5) with random displacements $\vec{t}_j \in$ {-0.2,+0.2} at coverage 0.045 after convergence of the mean values of the $\vec{s}_i$.

    Samples with greater values of $\vec{t}_j$ (not shown) indicate collective movement of adatoms. The clouds grow and often are drifting. We will return to the stability problems associated with random displacements of single adatoms below in section 4.3.

# 4. Superlattice Stability and Dynamic

- ### 4.1. Next Neighbor Model

To get an intuitive understanding of the stability of a hexagonal close packed superstructure we consider a simple next neighbor model consisting of a mobile adatom in the center of a fixed ring of 6 adatoms at distances a, interacting with $u(s)=u_0*s^{-3}$. Such potential is the limiting case of (2.3) for large s on an isotropic subatrate [6]. We can understand the ring of adatoms as being stabilized by the bulk of the superlattice.

The coordinates of this adatom configuration are (0,0) of the center adatom and (($a/2, a\sqrt{3}/2$),($a,0$),($a/2,-a\sqrt{3}/2$),($-a/2,-a\sqrt{3}/2$),($-a,0$),($-a/2, a\sqrt{3}/2$)) of the ring adatoms. The interaction $u_c(x,y)$ the center adatom sees when laterally dispaced (x,y) is

$$u_c(x, y) = u_0 * \left( \left((a/2-x)^2 + \left(\frac{\sqrt{3}}{2}a - y\right)^2\right)^{-\frac{3}{2}} + \left((a-x)^2 + (y)^2\right)^{-\frac{3}{2}} + \right.$$
$$\left((a/2-x)^2 + \left(\frac{\sqrt{3}}{2}a + y\right)^2\right)^{-\frac{3}{2}} + \left((a/2+x)^2 + \left(\frac{\sqrt{3}}{2}a + y\right)^2\right)^{-\frac{3}{2}} +$$
$$\left. \left((a+x)^2 + (y)^2\right)^{-\frac{3}{2}} + \left((a/2+x)^2 + \left(\frac{\sqrt{3}}{2}a - y\right)^2\right)^{-\frac{3}{2}} \right) \quad (4.1)$$

The second derivative of $u_c(x, y)$ at (0,0)

$$\partial_{x,x} u_c(0, 0) = \partial_{y,y} u_c(0, 0) = 27\, u_0 * a^{-5} \quad (4.2)$$

gives the restoring force constant of a 2-dimensional harmonic oscillator the center adatom forms in its neighboring ring. We conclude an increasingly rapid motion when the adatom distances shrink or the coverage increases.

Another interesting value describing the stability of this model is the escape potential of the center adatom, i.e. the difference between its potential in the midst of two ring adatoms and its rest potential

$$u_e(a) = u_c\left(0, \frac{\sqrt{3}}{2}\right) - u_c(0, 0) = \left(10 + 16\left(169\sqrt{7} + 49\sqrt{13}\right)\right) u_0 * a^{-3} \quad (4.3)$$

We conclude a strong increase of $u_e(a)$ with shinking adatom distance or increasing coverage. If we consider a threshold $u_t$, the size of the repulsive barrier between adatoms hindering formation of dimers, we even get an idea how adatom distance or coverage is related with the onset of nucleation. We separate 3 cases for the center adatom interaction $u_c$:
- $u_c < u_e$ and $u_c < u_t$ represents superlattice stability
- $u_c > u_e$ and $u_c < u_t$ represents superlattice instability against melting, i.e. dissolution of the hexagonal close packaging
- $u_c < u_e$ and $u_c > u_t$ represents superlattice instability against the formation of dimers or nucleation.

■ 4.2. Extended Neighbor Model

In a slightly less simple model a mobile central adatom is surrounded by fixed next neighbors up to the 9th n.n. of a hexagonal packed cluster. Such model is beyond a closed solution, so the acting potentials have been calculated numerically for different values of the superlattice constant a. Also the 2nd derivative at the origin has been calculated. To model the attractive short range potential, the cap (2.4) is omitted here and the anisotropic potential according to (2.3) is used.

Fig.2.a shows the resulting potentials as a function of the center adatom displacement with the superlattice constant a as parameter for the main directions <1-10> red and <-1-12 > blue broken.

The similarity with the model in section 4.1 is obvious though the values are higher due to the increased neighborhood. We note the potential decrease when the center adatom approaches its next neighbor due to the potential well in eq. (2.3). We note that the blue broken <-1-12> escape potential remains below the red <1-10> nucleation wall for a superlattice constant a>5, i.e. significant onset of nucleation for a<5. The potential increase when the center adatom has passed the mid of its next neighbors comes from the increasing repulsion of its second next neighbor. We should, however, not overestimate the relevance of this rigid neighbor model since the reaction of outer adatoms to the movement of the central adatom is ignored.

Fig.2.b shows the 2nd derivative of the potential as a function of the coverage for the main directions <1-10> red and <-1-12> blue broken.

u¨ measures the restoring force constant acting on the center adatom. The u¨$\sim a^{-5}$ relation of sect. 4.1 is visible here as u¨$\sim \theta^{5/2}$. This relation, however, is useful only for coverages up to 0.05.

- ### 4.3.Superlattice Dynamic

In an extended superlattice all adatoms are subject to potentials as discussed in sections 4.1 and 4.2 and thus oscillate in harmonic way as long as their displacements are limited. Ignoring for the moment the discrete structure of the substrate and the adatom locations we can utilize the concepts developed for crystal lattice dynamics as shown in textbooks like [10]. So collective vibrations, longitudinal and transverse waves with velocities proportional to the 2nd derivatives of the potential in the continuum limit can be expected like many other effects known from crystal lattice dynamics and acoustics. It is proposed to call the collective modes diffons in analogy to phonons. With increasing coverage diffusion would start with that of individual adatoms, gradually would develop towards collective diffusion and finally reach collective vibrations/diffons when the superlattice has formed and stabilized. Dimers in small quantities would act as impurities in such lattice.

Coming back to section 3.5 the covergence problems of eq. (2.5) with random displacement values $\vec{t}_j$>0 can be explained with $\vec{t}_j$ not reflecting collective modes.

# 5.Review of Experiments

- ### 5.1.Adatom Superstructures on Cu<111>

From the observations of Cu adatom superstructures on Cu<111> [3] over Ce/Cu<111> [4] to recent Fe/Cu<111> [5] superstructures this topic created much interest. It was reviewed recently [1]. Surface state mediated adatom interactions [11] were taken as explanation for those superstructures. In the Fe/Cu<111> case also strain relaxation effects were taken into account [5].

The superstructures show up at quite low temperatures and at adatom coverages below 0.05 ML. Their structure is hexagonal close packed with a lattice constant significantly greater than the substrate lattice constant.

- ### 5.2.Shortcomings of the current Explanation for Superlattices on Cu<111>

Starting from the Cu/Cu<111> experiments an oscillatory $s^{-2}$ type adatom-adatom interaction was derived from the adatom pair distribution using the zero coverage assumption of eq.(2.1) [3]. This assumption, however, is not valid in the 0.05 ML region for $s^{-3}$ interactions as the results of section 3 are showing where the $\theta$ term of (2.1) dominates. For an $s^{-2}$ interaction a zero coverage assumption is even more questionable. The oscillating $s^{-2}$ interaction thus was pretended and the conclusion of surface state mediated interactions needs to be reviewed.

The experiments also show an anisotropy of the adatom correlations which cannot be explained by an isotropic adatom-adatom interaction. An isotropic interaction would lead to a ring-type pair distribution and a distorted hexagonal close packaging if the coverage tends towards 0.05 ML [6].
In conclusion the isotropic surface state mediated interactions fail to explain the experimental results.

# 6.Discussion

- 6.1.Model Assumptions

Assumptions and approximations have been used for this model, discussed in [6] in more detail. We summarize here the most relevant ones:
- omitting all interactions but the elastic ones
- continuum model for the substrates instead of a lattice model, known to be inadequate for describing short range effects
- continuum model for the adatom distribution, again inadequate for describing short range effects
- the superposition approximation.
So conclusions always have to be drawn with caution.
If, however, the validity of eq. (2.1) is to be compared with the coverage $\theta$=0 assumption we are on the safe side: in the case of medium to long range interactions the pair distribution function depends not only on the 2-body forces between isolated adatoms but also on the interaction with the adatom medium which is expressed by the superposition approximation and is described in consequence by the Born-Green-Yvon equation. From the conclusions drawn in section 5.2 a review of the surface state interaction model in [3,4,5] and in related papers [12,13,14,15] seems appropriate.

- 6.2.Adatom Distribution on Cu <111>

The results of sections 3.1 and 3.4 together indicate the picture of a substrate aligned hexagonal dense packed adatom superlattice with voids when the coverage approaches its limiting value for the convergence of eq.(2.2). The number of dimers approaches a value of appr. 1/4 in this limiting case. When the coverage is below its limiting value the peaks in the pair distribution g(s) are less pronounced, the superlattice is less aligned, its distances are larger and the number of dimers is smaller. The results of sections 3.2 and 3.3 indicate a similar influence of the temperature: its decrease sharpens the superlattice, its increase degrades it. We note from section 3.5 the thermal movement of adatoms in their superlattice positions.

- 6.3.Superlattice Stability and Dynamic

The existence or stability of superlattices has several aspects.:
- Firstly convergence of (2.2) is an indicator of their stability in a certain coverage and temperature range. Figs. 1.c,d shows increasing pair distribution peaks with increasing coverage up to a certain limit. Figs. 1.e,f show degradation of such peaks with increasing temperature while Fig. 1.g,h show growth of such peaks i.e. increasing stability.
Fig. 1.i shows the meaning of stability: a well aligned grid. Fig. 1.j shows a stable configuration with adatoms oscillating around their stable average position.
- Secondly the simple models used in section 4 give an idea what stability means physically, the absence of nucleation and the confinement of adatoms in a potential well.
- Thirdly the Monte Carlo simulations with thermal movement in section 3.5 show the fragility of the lattice against random displacements. It is likely that collective displacement modes (diffons) would show a better stability.
- Last but not least the different methods above do not completely merge in distance and coverage values as expected when comparing effects of 2-body interactions with those including mean field effects.

Assessment of the stability range of superlattices is also difficult since nucleation and dissolution are competing and the continuum model used is unable to predict the true potentials acting on the short distances relevant for both effects. But more examples of superlattices should arrive when studying other surfaces of elastic strongly anisotropic materials.

The conclusions drawn in section 4.3 could reach beyond the adatom superlattice case. In cases where single adatoms are subject to repulsive interactions they may form liquid of freeze glass like structures. Adatoms may see more or less harmonic potentials and the theory of lattice dynamics for anharmonic effects, impurities and amorphous systems may be applied.

### 6.4. Adatom Diffusion

The conclusions drawn in section 4.3 on the gradual change of the character of diffusion have to be compared with analyses of collective and single particle diffusion [16,17] and with research on diffusion, especially on the dramatic decrease of the prefactor in certain cases [18]. The claim for collective modes / diffons goes beyond and of course needs experimental verification. Like inelastic scattering of neutrons in the case of bulk lattice dynamics, inelastic scattering on the surface would provide evidence and information about dispersion relations. The knowledge of dispersion relations in turn would allow comparisons with the results of section 4. The author would start evaluating the potential of 3He spin-echo measurements [19] for this purpose.

Neglecting the hopping between substrate lattice sites and the related activation energy is of course a strong assumption, valid only if the potential levels for hopping are smaller than those for adatom repulsion. The discrete nature of diffusion between substrate lattice sites will impose a structure to the dispersion relation which in turn may give insight to the acting potentials. Diffusion of adatoms is related to vibrations of the substrate lattice [20], so coupling between diffons and phonons could be a matter of further interest.

### 6.5. Open Questions

Though the calculations in this investigation can explain experiments at least qualitatively, the influence of the adatom kind is not fully answered. The only parameter considered is the strength of the stress magnitude parameter P (see section 2.2) which via the $\omega_p$ defines the magnitude of the interaction U and thus the temperature range T in which the scaled interaction $u_{12}=U_{12}/k_BT$ allows convergence of $g_{12}$ in eq. (2.1). In other words the higher the stress adatoms exert to the surface the higher is the temperature range in which superlattices can be expected.

The question what other parameters influence the existence or nonexistence of superlattices is beyond this simple continuum theory and needs consideration of other - especially short range - interactions.

So the present investigation can only reopen the race for valid answers in the search for adatom interactions explaining superlattices.

## 7. Summary

Substrate strain mediated adatom density distributions have been calculated for Cu>111> surfaces. Together with complementing Monte Carlo calculations they show the appearance of a hexagonal close packaged adatom superlattice in a coverage range up to 0.045 ML. This superlattice is not perfect but contains voids and dimers in the 0.045 ML region and is even less pronounced at coverages below. The influence of temperature has also been calculated, increasing temperatures lead to degradation of the superlattice, decreasing temperatures to a stabilization.

Stability and dynamic of the superstructure have been analyzed using simple neighborhood models. A harmonic motion of adatoms in potential wells was found limited by its escape from the rings and by nucleation.

Experimental results for adatoms on Cu<111> show a similar behavior but were explained by a surface state mediated interaction. It is argued that such explanation is due to a pretended oscillating $s^{-2}$ interaction. Furthermore it does not explain the alignment of the superstructure induced by the substrate anisotropy. Strain mediated interactions are proposed as alternative.

Adatom motion in superstructure potential wells and the analogy of lattice dynamics led to the conclusion that adatoms are moving in collective modes, a limiting case of diffusion. Inelastic scattering is proposed as a means to measure such modes.

## Acknowledgement

Many thanks to my wife for patience and support.

# Appendix

### ■ Figures

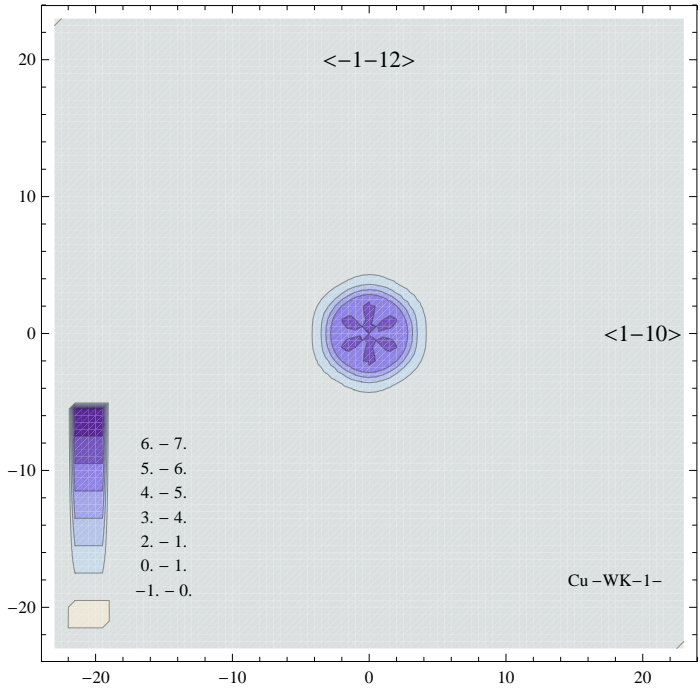

Fig.1.a: u(s) contour plot, dark colors represent high values, light colors represent low values, contours changing in 1- steps

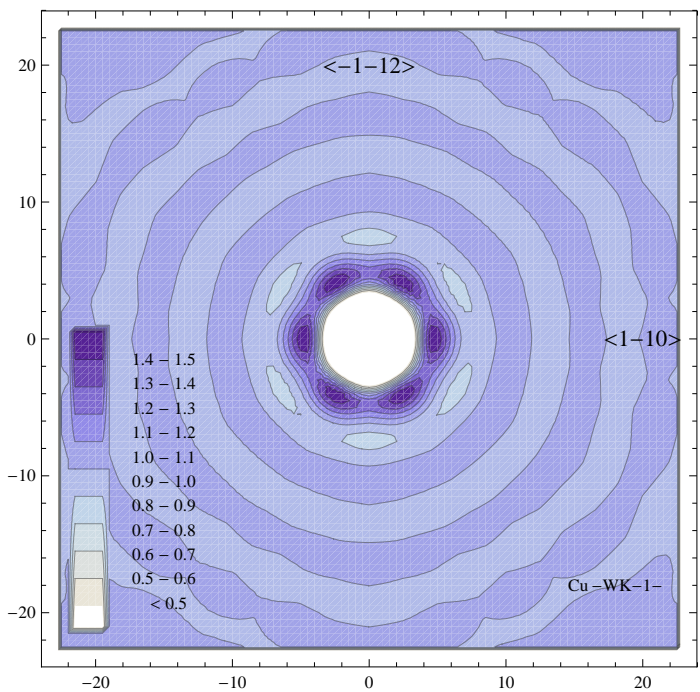

Fig.1.b: g(s) contour plot for the limiting coverage $\theta_c$=0.045, contours changing in 0.1 steps

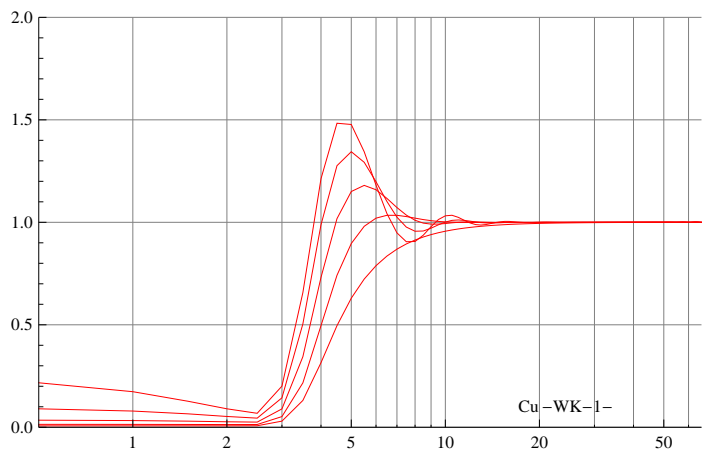

Fig.1.c: semi-logarithmic $g_{<1-10>}(s,\theta)$ plot for s ∥ <1-10>, showing its evolution with increasing coverage $\theta$ in 0.009 steps

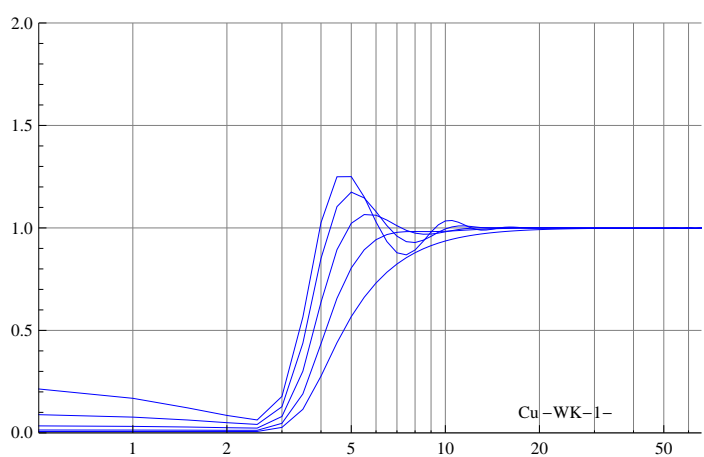

Fig.1.d: semi-logarithmic $g_{<-1-12>}(s,\theta)$ plot for s ∥ <-1-12>, showing its evolution with increasing coverage $\theta$ in 0.009 steps

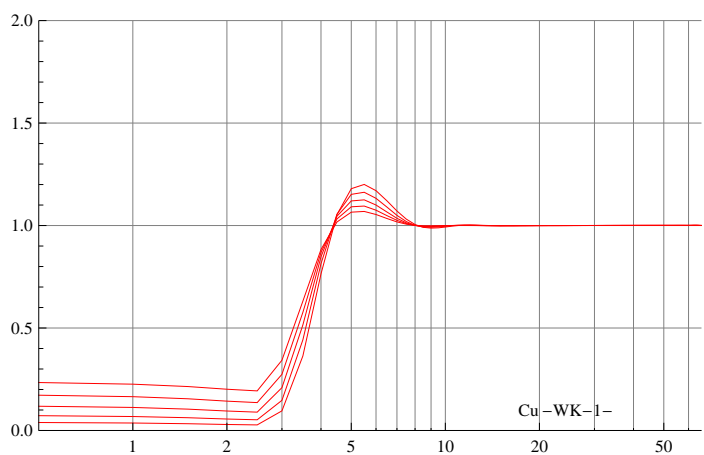

Fig.1.e: semi-logarithmic $g_{<1-10>}(s,\theta)$ plot for s ∥ <1-10>, showing its evolution with temperature T increasing from 1 to 2.2

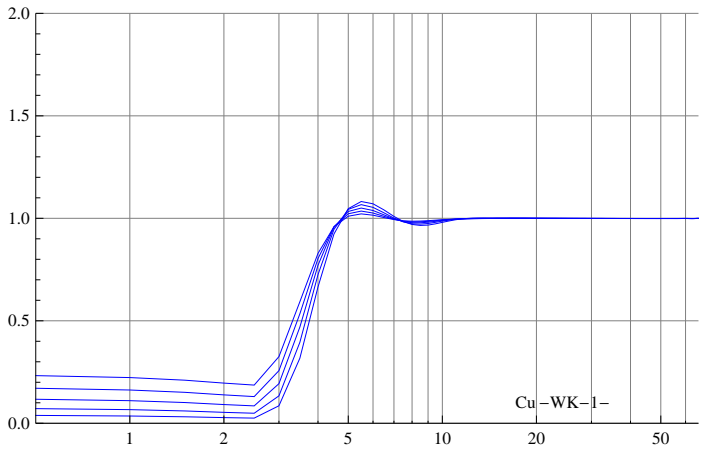

Fig.1.f: semi-logarithmic $g_{<-1-12>}(s,\theta)$ plot for s ∥ <-1-12>, showing its evolution with temperature T increasing from 1 to 2.2

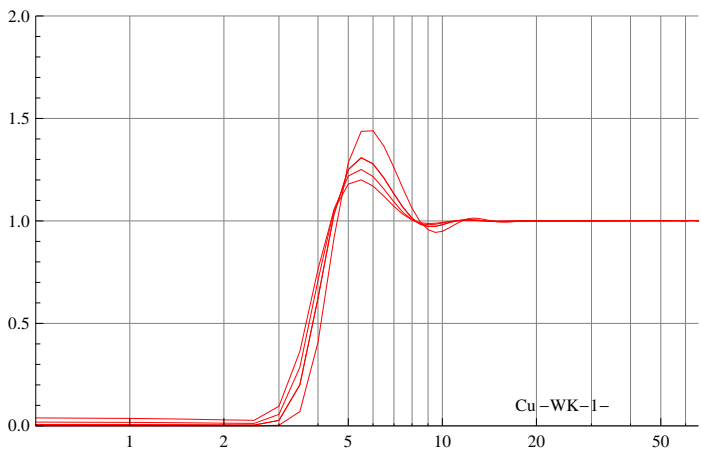

Fig.1.g: semi-logarithmic $g_{<1-10>}(s,\theta)$ plot for s ∥ <1-10>, showing its evolution with temperature T decreasing from 1 to 0.44

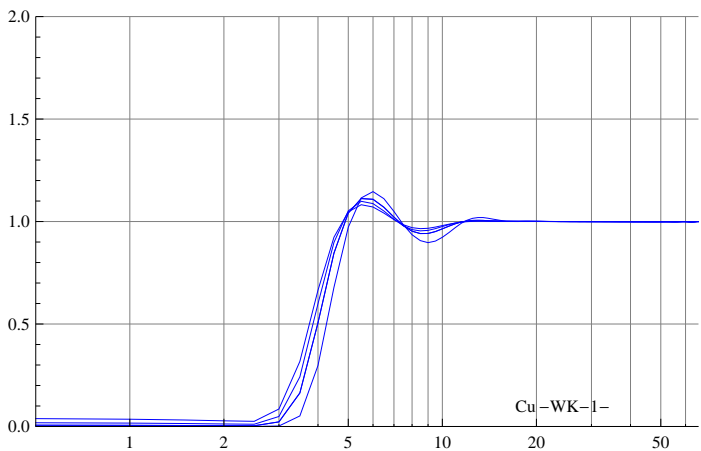

Fig.1.h: semi-logarithmic $g_{<-1-12>}(s,\theta)$ plot for s ∥ <-1-12>, showing its evolution with temperature T decreasingfrom 1 to 0.44

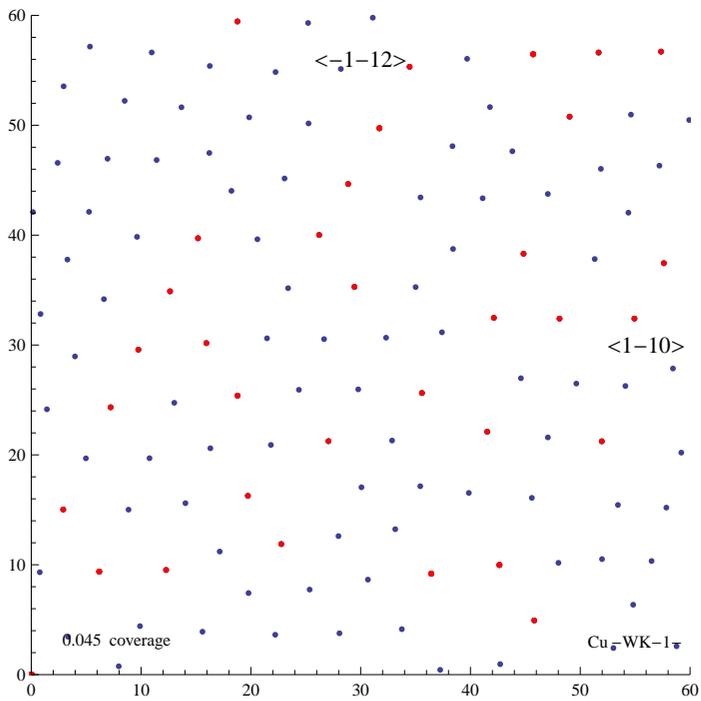

Fig.1.i: adatom position sample from a Monte Carlo simulation at coverage $\theta$=0.045. Dimers or clustered objects with a distance $< s_w$ are depicted in red, others in blue color.

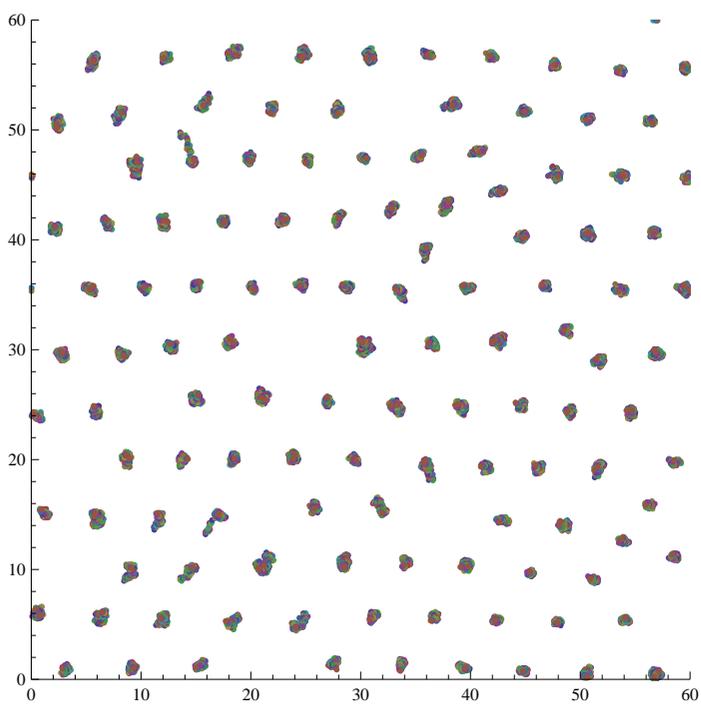

Fig.1.j: adatom trajectory sample from a Monte Carlo simulation at coverage 0.045

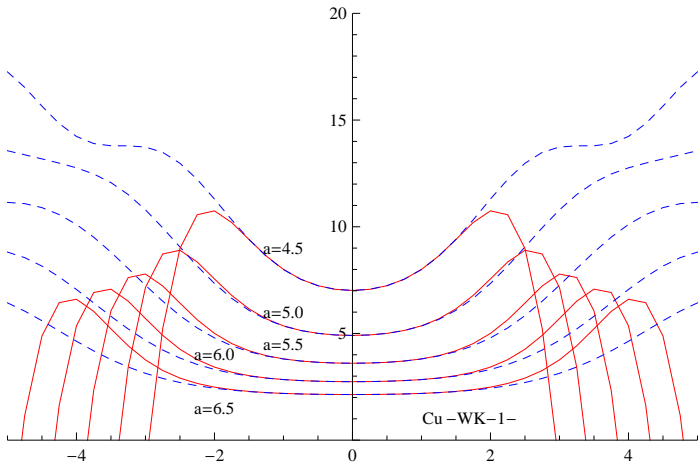

Fig.2.a: Extended neighbor model potentials as a function of the center adatom displacement with the superlattice constant a as parameter for the main directions <1-10> red and <-1-12 > blue broken.

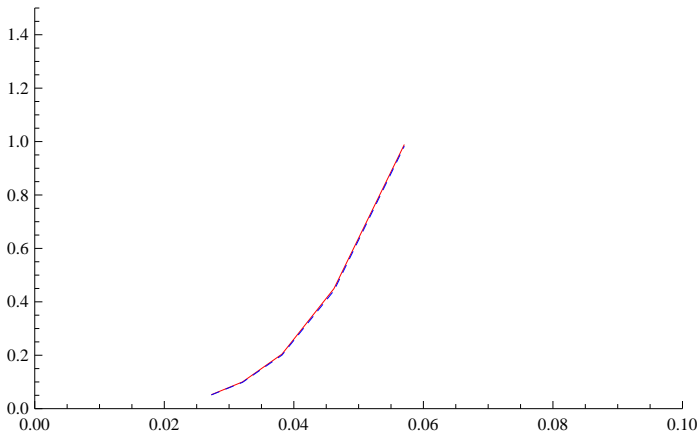

Fig.2.b 2nd derivative of the extended neighbor potential as a function of the coverage for the main directions <1-10> red and <-1-12> blue broken.